\documentclass[letterpaper, amsfonts, amssymb, amsmath, reprint, showkeys, nofootinbib, twoside,superscriptaddress]{revtex4-1}
\usepackage[english]{babel}
\usepackage[utf8]{inputenc}
\usepackage[colorinlistoftodos, color=green!40, prependcaption]{todonotes}
\usepackage[pdftex]{hyperref} 
\newcommand{\papertitle}{Coupling of Light and Mechanics in a Photonic Crystal Waveguide}

\newcommand{\linkcolor}{magenta}%
\hypersetup{colorlinks=true, %
  linkcolor=\linkcolor, %
  citecolor=\linkcolor, %
  filecolor=\linkcolor, %
  urlcolor=\linkcolor, %
  pdftitle=\papertitle,
  pdfauthor={J.-B. Béguin, Z. Qin, X. Luan, H. J. Kimble}, %
  pdfsubject={}, %
  pdfkeywords={} %
}

\usepackage{siunitx}
\DeclareSIUnit\torr{torr}

\usepackage{caption,subcaption}
\usepackage{ragged2e}
\DeclareCaptionJustification{justified}{\justifying}
\begin{document}
\title{\papertitle}

\author{J.-B. B\'eguin}
    \affiliation{Norman Bridge Laboratory of Physics MC12-33, California Institute of Technology, Pasadena, CA 91125, USA}
    \affiliation{Present address: Niels Bohr Institute, University of Copenhagen, Blegdamsvej 17, 2100 Copenhagen, Denmark}

\author{Z. Qin}    
    \affiliation{Norman Bridge Laboratory of Physics MC12-33, California Institute of Technology, Pasadena, CA 91125, USA}
    \affiliation{Other affiliation: State Key Laboratory of Quantum Optics and Quantum Optics Devices, Institute of Opto-Electronics, Shanxi University, Taiyuan 030006, China}
    
\author{X. Luan}
    \affiliation{Norman Bridge Laboratory of Physics MC12-33, California Institute of Technology, Pasadena, CA 91125, USA}

\author{H. J. Kimble}
    \email[Correspondence email address: ]{hjkimble@caltech.edu}
    \affiliation{Norman Bridge Laboratory of Physics MC12-33, California Institute of Technology, Pasadena, CA 91125, USA}

\date{\today} 

\begin{abstract}
Observations of thermally driven transverse vibration of a photonic crystal waveguide (PCW) are reported. The PCW consists of two parallel nanobeams with a \SI{240}{\nano\meter} vacuum gap between the beams. Models are developed and validated for the transduction of beam motion to phase and amplitude modulation of a weak optical probe propagating in a guided mode (GM) of the PCW for probe frequencies far from and near to the dielectric band edge. Since our PCW has been designed for near-field atom trapping, this research provides a foundation for evaluating possible deleterious effects of thermal motion on optical atomic traps near the surfaces of PCWs. Longer term goals are to achieve strong atom-mediated links between individual phonons of vibration and single photons propagating in the GMs of the PCW, thereby enabling opto-mechanics at the quantum level with atoms, photons, and phonons. The experiments and models reported here provide a basis for assessing such goals, including sensing mechanical motion at the Standard Quantum Limit (SQL).
\end{abstract}


\maketitle

Recent decades have seen tremendous advances in the ability to prepare and control the quantum states of atoms, atom-like systems in the solid state, and optical fields in cavities and free space. However, the integration of these diverse elements to achieve efficient quantum information processing still faces diverse challenges, including the wide range of highly dissimilar physical systems (e.g., atoms, ions, solid-state defects, quantum dots) that could be utilized to realize heterogeneous systems for quantum logic, memory, and long-range coupling. Each of these systems has unique advantages, but they are disparate in their frequencies, their spatial modes, and the fields to which they couple. For example, the electronic degrees of freedom in atoms and atom-like defects typically respond at optical frequencies, while their spin degrees of freedom, which are suitable for long-term storage of quantum states, respond to microwave or radio frequencies. On the other hand, the transmission of quantum information over long distances at room temperature requires the use of telecom-band photons in single-mode optical fibers. 

Beginning with the pioneering work in Refs. \cite{braginsky01,rokhsari05}, mechanical systems have now been recognized as broadly applicable means for overcoming these disparities and transferring quantum states between different quantum degrees of freedom \cite{hammerer09,rabl10,rabl09,hammerer09b}. This is because mechanical systems \cite{aspelmeyer14} can be engineered to couple efficiently and coherently to many different systems and can possess very low damping, particularly when operated at cryogenic temperatures. To date, quantum effects have been observed in mechanical systems coupled to superconducting qubits (via piezoelectric coupling) \cite{oconnell10}, optical photons \cite{chan11,purdy13,safavi-naeini12,brooks12,safavi-naeini13,purdy13b}, and microwave photons \cite{teufel11,palomaki13}. Efficient coupling has also been demonstrated between mechanical oscillators and spins in various solid-state systems, although to date the mechanical components of these devices have operated in the classical regime \cite{rugar04,hong12,kolkowitz12,arcizet11,macquarrie13,teissier14,ovartchaiyapong14}.

In this manuscript we describe nascent efforts to utilize strong coupling of atoms, photons, and phonons in nanophotonic PCWs to create a new generation of capabilities for quantum science and technology. Our long-term goal is to use optomechanical systems operating in the quantum regime to realize controllable, coherent coupling between isolated, few-state quantum systems. In our case, the system will consist of atoms trapped along a photonic crystal waveguide (PCW) that interact strongly with photons propagating in the guided modes (GMs) of the PCW \cite{chang2018colloquium}. The mechanical structure of the PCW in turn supports phonons in its various eigenmodes of motion. While much has been achieved in theory and experiment for strong coupling of atoms and photons in nano-photonics, much less has been achieved (or even investigated) for the optical coupling of motion and light in the quantum regime for devices such as described in Refs. \cite{chang2018colloquium,lodahl2017chiral}. 

\begin{figure*}[t]
\centering
\includegraphics[width=1.0\linewidth]{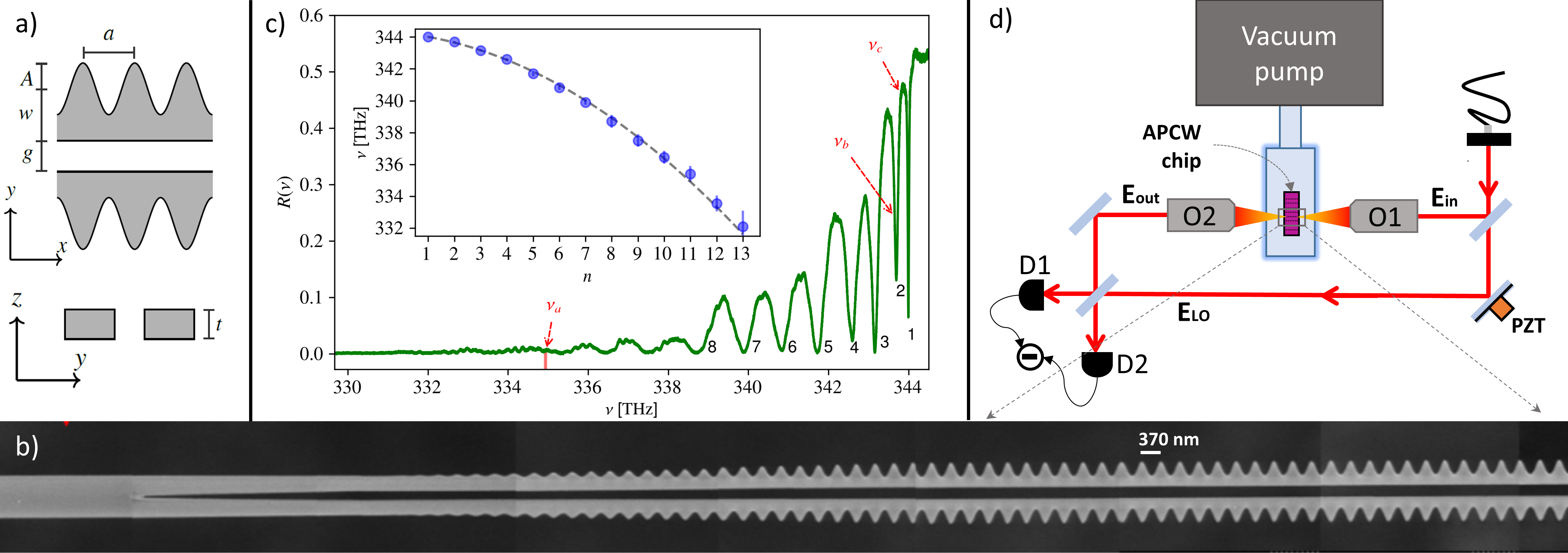}
\caption{Details of the alligator photonic crystal waveguide (APCW) and the setup for our experiments \cite{yu2014nanowire,hood2016atom,yuthesis,mcclungthesis}. a) Drawing giving the dimensions of the various components of the APCW in gray. The unit cell spacing $a= \SI{370}{\nano\meter}$, the vacuum gap $g=\SI{238}{\nano\meter}$, and the Silicon Nitride thickness $t = \SI{200}{\nano\meter}$. The outer beams have modulation amplitude $A=\SI{120}{\nano\meter}$ and width $w=\SI{280}{\nano\meter}$.  b) An SEM image of the left half of the APCW showing (from left to right) a single unstructured rectangular waveguide that splits at a Y-junction into two parallel waveguides each of which is gradually modulated in width to finally match the $A,w$ values of the APCW itself which extends $150$ unit cells to the right along $x$ before tapering to a second Y-junction and a uniform rectangular beam. The entire structure is suspended in vacuum by transverse tethers connected to supporting side rails (not shown) \cite{yu2014nanowire,yuthesis,mcclungthesis}. c) Reflection spectrum $R(\nu)$ for the APCW displays a series of low finesse cavity-like resonances for reflections from the input tapers and APCW near the dielectric band edge at $\SI{344}{\tera\hertz}$. The inset plots frequencies $\nu_n$ for successive cavity resonances $n=1,2,...$ near the dielectric band edge. d) Simplified diagram for measurements of mechanical modes of the APCW by way of transmission spectra $T(\nu)$ either by direct detection of beam $E_{out}(\nu)$ alone at photodetector $D_1$ or $D_2$, or via balanced homodyne detection of the signal beam $E_{out}(\nu)$ combined with the local oscillator beam $E_{LO}(\nu)$ at photodetectors $D_1$ and $D_2$.}
\label{fig:setup}
\vspace{-10pt}
\end{figure*}

A longstanding challenge for this work is to achieve the integration of ultracold atoms with nanophotonic devices. If this challenge were overcome, quantum motion could be harnessed to investigate enhanced nonlinear atom-light interactions with single and multiple atoms. New quantum phases \cite{manzoni17}, novel mechanisms for controlling atoms near dielectric objects \cite{chang2013trapping}, and strong atom-photon-phonon coupling \cite{hammerer09b} could be realized in the laboratory. Although difficult, this approach potentially benefits from several advantages when compared to conventional optomechanics, including (a) the extreme region of parameter space that atomic systems occupy (such as low mass and high mechanical Q factors), (b) the exquisite level of control and configurability of atomic systems, and (c) the pre-existing quantum functionality of atoms, including internal states with very long coherence times.

Of course, many spectacular advances of atomic physics already build upon these features \cite{blatt08,haffner08,monroe10}. On one hand, experiments with linear arrays of trapped ions achieve coherent control over phonons interacting with the ions' internal states as pseudo spins. Goals that are very challenging for quantum optomechanics with nano- and micro-scopic masses, such as phonon-mediated entanglement of remote oscillators and single-phonon strong coupling, are routinely implemented with trapped ions. On the other hand, cavity QED with neutral atoms produces strong interactions between single photons and the internal states of single atoms or ensembles, leading to demonstrations of state mapping and atom-photon entanglement \cite{rempe15}.

What is missing thus far, and what motivates the initial steps described here, is a strong atom-mediated link between individual photons and phonons, to enable optomechanics at the quantum level. Initial steps described here include 1) observation and characterization of the low frequency, mechanical eigenmodes of an alligator photonic crystal waveguide (APCW) and 2) the development of theoretical models that are validated in the nontraditional regime in which our system works \cite{shelby85b,shelby85}, namely, well localized mechanical modes, but non-localized propagating photons both far from and near to the band edges of PCWs.

\section{The alligator photonic crystal waveguide}

Figure \ref{fig:setup} provides an overview of the APCW utilized in our experiments with details related to device fabrication and characterization provided in Refs. \cite{yu2014nanowire,hood2016atom,yuthesis,mcclungthesis}. The photonic crystal itself is formed by external sinusoidal modulation of two parallel nano-beams to create a photonic bandgap for TE modes with polarization predominantly along $y$ in Figure \ref{fig:setup}(a). The TE band edges have frequencies near the D1 and D2 transitions in atomic Cesium (Cs). Calculated and measured dispersion relations for such devices are presented in Ref \cite{hood2016atom} where good quantitative agreement is found. Here, we focus on coupling of light and motion for TE modes of the APCW. TM modes of the APCW near the TE band edges resemble the guided modes of an unstructured waveguide.

As shown by the SEM image in Fig.~\ref{fig:setup}(b), the APCW is connected to single-beam waveguides on both end and thereby freely suspended in the center of a 2 mm wide window in a Silicon chip.  Well beyond the field of view in Fig.~\ref{fig:setup}(b), a series of tethers are attached transversely to the single-beam waveguides along $\pm y$ to anchor the waveguides to two side rails that run parallel to the $x$ axis of the device to provide thermal anchoring and mechanical support, with the coordinate system defined in Fig.~\ref{fig:setup}(a). Important for our current investigation, the single-beam waveguides and the APCW itself are held in tension with $T \simeq \SI{800}{\mega\pascal}$.

Light is coupled into and out of TE guided modes of the APCW by a free-space coupling scheme that eliminates optical fibers within the vacuum envelope~\cite{beguin20,luan2020}. An example of a reflection spectrum $R(\nu)$ is given Fig. \ref{fig:setup}(c), which is acquired by way of light launched from and recollected by the microscope objective \textit{O1} shown in Fig. \ref{fig:setup}(d). Objectives \textit{O1, O2} are mode-matched to the fields to/from the terminating ends of the waveguide resulting in overall throughput efficiency $T \simeq 0.50$ from input objective \textit{O1} through the device with the APCW to output objective \textit{O2} for the experiments described here. The silicon chip itself contains a set of APCWs and is affixed to a small glass optical table inside a fused silica vacuum cell by way of silicate bonding \cite{beguin20,luan2020}.

\section{Observations of modulation spectra}

With reference to Fig.~\ref{fig:setup}(d), we have recorded spectra $\Phi(\nu,f,\theta)$ for the difference of photocurrents from detectors $D_1,D_2$ for light transmitted through an APCW for various probe frequencies $\nu$ below the frequency $\nu_{BE}\simeq\SI{344}{\tera\hertz}$ of the dielectric band edge. Here we employ a balanced homodyne scheme with $E_{in}$ and $E_{LO}$ having identical optical frequency $\nu$ and each absent radio frequency modulation $f$ save that from propagation in the APCW. With free-space coupling to guided modes of the APCW, homodyne fringe visibility up to $\sim 0.95$ is obtained.

\begin{figure}[t]
\centering
\includegraphics[width=\linewidth]{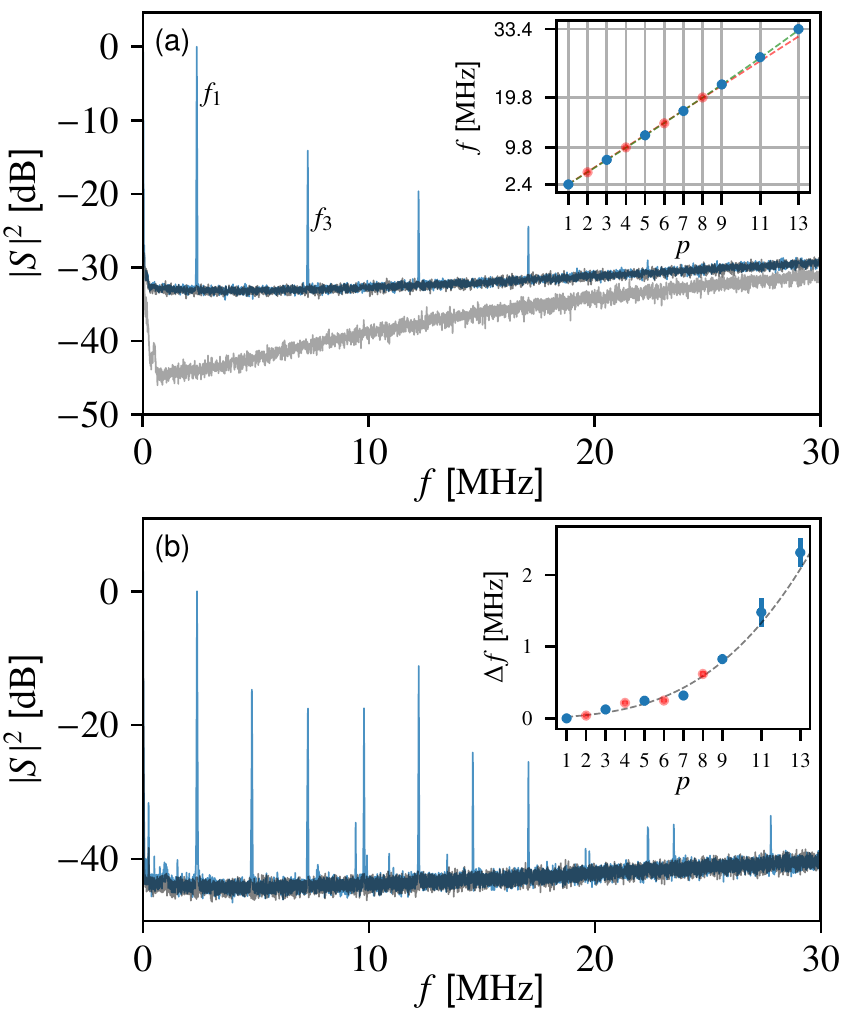}
\caption{Measured vibration spectra with electrical spectrum analyzer (bandwidth \SI{10}{\kilo\hertz}) at wavelength $895.00$ nm (a)  and $872.40$  nm (b), respectively. Electronic noise floor of the homodyne detector is shown in gray in part (a). Inset in (a) plots the frequencies $f_p$ of odd quasi-harmonics peaks (blue dots) and even quasi-harmonics peaks (red dots) of $f_1$. Linear fit (dashed red curve) and complete fit of $\tilde{f}_p$ (dashed green curve) are also shown. Inset in (b) plots the measured frequency difference $\Delta f$ (blue and red dots) and theoretical fit $\Delta \tilde{f}$ (dashed grey curve).}
\label{fig:spectra}
\end{figure}

\begin{figure}[t]
\centering
\includegraphics[width=\linewidth]{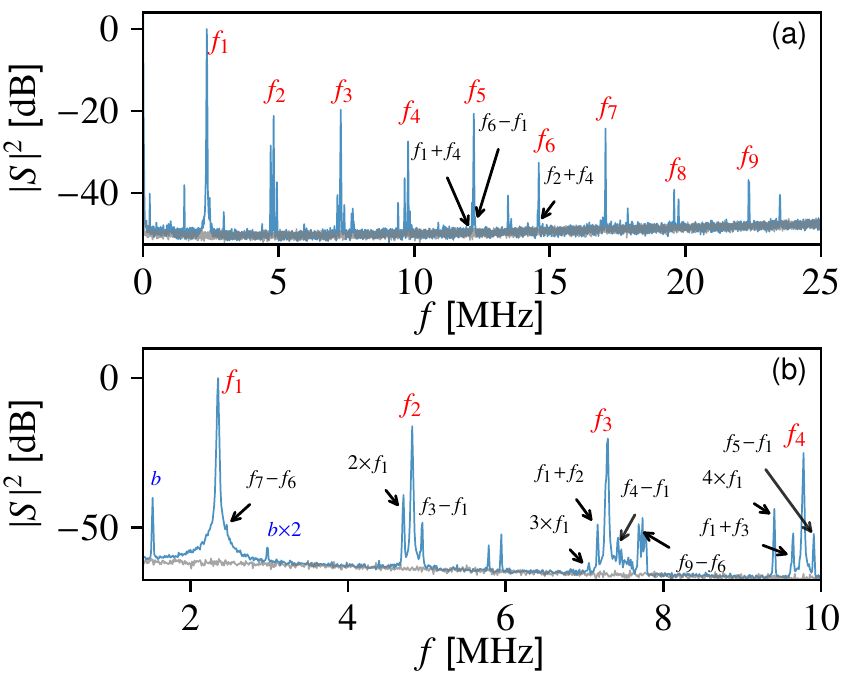}
\caption{Measured vibration spectra near the band edge at $\SI{872.04}{\nano\meter}$ ($\nu_c$) with spans of 25 MHz (a) and 10 MHz (b). Except for the dominant peaks appearing in Fig.~\ref{fig:spectra} (i.e, approximate integer harmonics of $f_1$ labelled in red), sums and differences of the dominant quasi-harmonics frequency components are also observed. Peak labeled as $b$ is due to unbalanced input laser light noise.}
\label{fig:spectraBE}
\end{figure}

Measurement results for $\Phi(\nu,f,\theta)$ are displayed in Figures \ref{fig:spectra} and \ref{fig:spectraBE} for three optical frequencies $\{\nu_a,\nu_b,\nu_c\}= \{334.96,343.64,\SI{343.78}{\tera\hertz}\}$ (i.e., wavelengths $\{895.00,872.40,\SI{872.04}{\nano\meter}\}$) moving from far below to near the dielectric band edge, as marked by red arrows in Figure \ref{fig:setup}c. The spectra display a series of narrow peaks and are of increasing complexity as the band edge is approached. All spectra are taken for a weak probe beam $E_{out}(\nu)$ with power $P_{out} \sim \SI{10}{\micro\watt}$, while $P_{LO} \simeq \SI{0.5}{\milli\watt}$. The phase offset $\theta$ between $E_{in}$ and $E_{LO}$ is set to maximize the observed spectral peaks whose frequencies $f$ exhibit only small shifts with changes in $P_{out}$, as illustrated in Figure \ref{fig:S1} in \cite{SM}. In vacuum ($\sim \SI{1e-10}{\torr}$) and at room temperature, the quality factor for the lowest peak at $f_1 \simeq \SI{2.4}{\mega\hertz}$ is $Q \simeq \SI{1e5}{}$. This value is compatible with the numerically predicted increase of the intrinsic $Q_{\text{int}} \simeq \SI{8.4e3}{}$ from the high material pre-stress for $\SI{200}{\nano\meter}$ thin SiN beams \cite{villanueva14}.

An important feature of the spectra in Figure \ref{fig:spectra}(a) is that peaks beyond $f_1$ occur at frequencies that are approximately odd harmonics of $f_1$, with $f_j \simeq j\times f_1$ for $j=1,3,5,...$. By contrast in Figure \ref{fig:spectra}(b), the largest peaks double in number with now the presence of \textit{even} harmonics of the fundamental frequency $f_1$ in addition to the \textit{odd} harmonics from Figure \ref{fig:spectra}(a). As shown by the inset in Figure \ref{fig:spectra}(a), the dispersion relation is approximately linear with frequencies $f_p \simeq p\times f_1$, where $p=1,2,3,...$.

Further understanding emerges if we consider higher accuracy for the frequencies $f_p$  and examine the measured frequency differences $\Delta f=\{f_p - p f_1\}$ as in the inset of Figure \ref{fig:spectra}(b). Also plotted as the dashed line is the theoretical prediction for the mechanical frequency differences $\Delta \tilde{f} = \{\tilde{f}_p - p\tilde{f}_1\}$ of a long, narrow, and thin beam, which is supported at hinged ends. For this model, the mechanical resonances are \citep{Hocke2014}
\begin{align}
\tilde{f}_p = \frac{p^2\pi}{2L^2}\sqrt{\frac{EI}{\rho A}+\frac{\sigma L^2}{\rho \pi^2 p^2}}, \label{opto:eq1}
\end{align}
where $p$ is the integer mode index, $E$ the Young's modulus, $I$ the moment of inertia, $A$ the cross sectional beam area, $L$ the beam length, $\rho$ the mass density, and $\sigma$ the beam stress. 

Our APCW and connecting nano-beams are fabricated from SiN with high-tensile stress $\sigma \simeq \SI{800}{\mega\pascal}$ \cite{Yu14,yuthesis}. Together with the largely 1D geometry of the APCW (large aspect ratio of transverse to longitudinal dimension), the contribution of the bending term in $\eqref{opto:eq1}$ can be neglected for the lowest order modes such that $\tilde{f}_p \simeq (p/2L)\sqrt{\sigma/\rho}$, giving rise to a close approximation of the linear dispersion of a tensioned string as in the inset to Figure \ref{fig:spectra}(a). However, higher order modes have a clear quadratic contribution from the bending term that is evident in the inset to 
Figure \ref{fig:spectra}(b).

\begin{figure*}[t]
\centering
\includegraphics[width=1.0\linewidth]{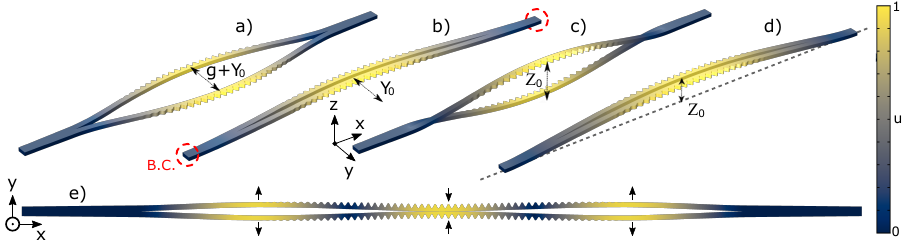}
\caption{Mechanical modes of the APCW structure illustrated with a reduced geometry. Four types of eigenmodes are shown (a) $Y_1^A$, (b) $Y_1^S$, (c) $Z_1^A$, (d) $Z_1^S$; where the total number of APCW unit cells is $N=10$, the total number of taper cells $N_t = 15$ and the length of a single Y-split junction $L_Y = \SI{6}{\micro\meter}$. (e) shows a higher-order mode $Y_3^{A}$ for a longer structure with $N=30,N_t =30, L_Y = \SI{12}{\micro\meter}$. B.C. means two end-clamped boundary conditions. Arbitrary displacement amplitude scales were chosen for illustration purposes (see main text).}
\label{fig:modes}
\end{figure*}

In terms of absolute agreement between measured and predicted frequencies for the spectra in Fig. \ref{fig:spectra}, from Eq. \ref{opto:eq1} we calculate a fundamental frequency $\tilde{f}_1 = \SI{2.37\pm0.3}{\mega\hertz}$ from the total length $L = 180\times\SI{0.37}{\micro\meter}+2\times\SI{20}{\micro\meter} = \SI{107\pm10}{\micro\meter}$, the manufacturer's quoted tensile stress $\sigma = \SI{800\pm50}{\mega\pascal}$, and the mass density for LPCVD (stochiometric) Silicon Nitride \citep{Pierson}, $\rho_{\text{SiN}} = \SI{3180}{\kilo\gram\per\cubic\meter}$. For the length $L$, we consider the $150$ unit cells of the actual PCW region, plus the 30 tapered cells on each end, and finally the length from the beginning of the Y-split junction which separates the two corrugated beams. The devices are designed for small stress relaxation from that of the original SiN on Silicon chip \cite{yuthesis}. The predicted $\tilde{f}_1$ is close to the measured frequency $f_1 = \SI{2.3844}{\mega\hertz}$.

While the frequencies of the largest peaks in Figure \ref{fig:spectra} are well-described by Eq. \ref{opto:eq1}, the complexity of the spectra increases as the band edge is approached with the appearance of many small satellite peaks as in Figure \ref{fig:spectraBE} for $\nu_c= \SI{343.78}{\tera\hertz}$ (i.e., wavelength $\lambda_c = \SI{872.04}{\nano\meter}$).

After labeling for clarity the dominant even and odd quasi-harmonics that also appear in Figure \ref{fig:spectra}, we clearly observe a secondary series of integer harmonics in  Figure \ref{fig:spectraBE}, such as the second, third and fourth harmonics of the lowest frequency $f_1$. The majority of the remaining peaks have frequencies which coincide with sums and differences of the main quasi-harmonics frequency components $f_p$. Other peaks (e.g., at \SI{1.5}{\mega\hertz}) originate from unbalanced input laser light noise.

\section{Mechanical modes of the APCW}

From measurements as in Figures \ref{fig:spectra} and \ref{fig:spectraBE} in hand and some understanding of the dispersion relation for the observed mechanical modes of the APCW, we turn next to more detailed characterization by way of numerical simulation. Principal goals are 1) to determine the mechanical eigenfunctions (and not just eigenfrequencies) associated with the observed modulation spectra and 2) to investigate the transduction mechanisms that convert mechanical motion of the various eigenfunctions to modulation of our probe beam. Beyond numerics to find the mechanical eigenmodes, we will present simple models to describe the transduction of mechanical motion to light modulation for various regimes far from and near to a band edge of the APCW. Quantitative numerical evaluation of the opto-mechanical coupling coefficient $G_{\nu}$ and eigenmodes for the full APCW structure will be presented in Section $5$.

Figure \ref{fig:modes} shows the fundamental mechanical modes of a small APCW structure obtained via numerical solution of the elastic equations. For clarity, we illustrate with a reduced geometry due to the large aspect ratio of our structure. The top panels represents the 3D deformed geometry as prescribed by the displacement vector field associated to each of the mechanical eigenmodes, with an arbitrary choice of mechanical energy. The displacement $u$ normalized to its maximum value $u_{max}$ is indicated by the colormap. The bottom panel displays a higher-order anti-symmetric mode with $f_{3}^{y,A} \sim 3 f_{1}^{y,A}$ in the x-y plane for a longer structure

The design of the relatively long Y-junction arises from the need for efficient (i.e., adiabatic) conversion of the light guided from the single waveguide into the mode of the double-beam photonic crystal. While it does not represent a sharp boundary for the mechanics (please refer to refs.~\cite{yuthesis,mcclungthesis} for details of full suspended structure with anchoring tethers), it does impose a symmetric termination geometry for both patterned beams. For the choice of effective two end-clamped boundary conditions, the four types of eigenmodes consist of two pairs of symmetric $S$ and antisymmetric $A$ oscillation, one pair with motion predominantly along $y$, which we denote by $Y_p^{A},Y_p^{S}$, and the other with motion mainly along $z$, denoted by $Z_p^{A},Z_p^{S}$ and labelled by the mode number $p=1,2,3,..$. For the actual full APCW structure, the eigenfrequencies for the fundamental $p=1$ modes are in the ratio $f_{1}^{y,A}, f_{1}^{y,S} f_{1}^{z,A}, f_{1}^{z,S} = 1,0.77,0.98,0.74$.

While the modes in Figure \ref{fig:modes} correspond to the mode families with lowest eigenfrequencies, at higher frequency other type of beam motion with mixed $y{-}z$ displacements appear. Also, as discussed in the conclusion, the APCW is a 1D phononic crystal. The eigenmodes shown in Figure \ref{fig:modes} correspond roughly to those of two weakly coupled nanobeam oscillators. Regarding the accuracy of the choice of boundary condition, we note that the mechanical properties of the differential modes are little impacted by the length of the single beam beyond the merging point of the junction.

\section{Mapping motion to optical modulation}

\subsection{Optical frequencies far from a band edge}

A simple model for the transduction of motion of the APCW nano-beams into optical modulation explains some of the key observations from the previous sections. First of all, for a fixed GM frequency $\omega$ input to the APCW, each \textit{mechanical} eigenmode adiabatically modifies the band structure of the APCW and thereby the \textit{optical} dispersion relation $k_x(\omega)$ for GM propagation along $x$ with frequency $\omega$ relative to the case with no displacement from equilibrium. In our original designs of the APCW, we undertook extensive numerical simulations of the band structure for variations of all the dimensions shown in Fig.~\ref{fig:setup}(a) Refs.~\cite{yuthesis,mcclungthesis,hood2016atom}. Guided by these earlier investigations, we deduce that the largest change in band structure with low frequency motion as in Fig. \ref{fig:modes} arises from variation of the gap width $g$ from displacements $\pm \delta y/2$ for the antisymmetric eigenmode $Y_p^{A}$ illustrated in Figure \ref{fig:modes} (a).

As suggested by Eq.~\ref{opto:eq1}, we then consider a $1D$ string model with $Y_p^{A}(x)$ describing $y$ displacement at each point along $x$, namely $Y_p^{A}(x)=Y_{0,p} \sin(\beta_p x)$, with maximum $y$ displacement $Y_{0,p}$. Here, $\beta_p$ is the mechanical wave vector with $Y_p^{A}(x)$ subject to boundary conditions, which in the simplest case are $Y_p^{A}(x=0)=0= Y_p^{A}(L)$ with then eigenvalues $\beta_p = p\pi/L$ for $p=1,2,3,...$. Again, $Y_p^{A}(x)$ denotes the mechanical eigenmode in Fig.~\ref{fig:modes}(a) and represents antisymmetric $y$ displacements of each nanobeam, with one beam of the APCW having displacement from equilibrium $\pm \delta y = \pm Y_0/2$ and the opposing beam with phase-coherent displacement  $\mp \delta y = \mp Y_0/2$, leading to a cyclic variation of the total gap width $g \rightarrow g + Y_0 \rightarrow g \rightarrow g - Y_0 \rightarrow g $ as described by $Y_p^{A}(x)$ along the $x$-axis of the APCW. For small $y$ displacements and fixed frequency $\omega$ far from the band edge, we can then expand the dispersion relation to find $k_x(\omega, y) \simeq k_x(\omega, 0) + \delta k_x(\omega, y)$, where $\delta k_x(\omega, y) = \xi(\omega)\times y$, with $\xi(\omega)=(\frac{dk_x(\omega)}{dy})$.

Since $y$ displacements vary along $x$ as described by the particular mechanical eigenmode $Y_p^{A}(x)$, $\delta k_x$ will also vary along $x$. The differential phase shift due to a mechanical eigenmode for propagation of an optical GM from input to output of the APCW is then given by (in our simple model) $\Phi_p(L) = \int_0^L  \delta k_x(\omega, Y_p^{A}(x))dx = \int_0^L \xi(\omega) Y_p^{A}(x) dx = 2L \xi(\omega)Y_{0}/p\pi$ for $p$ odd, and $\Phi_p(L) = 0$ for $p$ even. Here, $\Phi_p(L)$ is the differential phase shift between optical propagation through the APCW with and without mechanical motion (i.e., $Y_{0,p} \neq0$ and $Y_{0,p}=0$).

When driven by thermal Langevin forces, the mechanical mode $Y_p^{A}(x)$ oscillates principally along $y$ at frequency $f_{p}^{y,A}$ with rms amplitude $\langle Y_{0,p}^2 \rangle^{1/2}$, where $\langle Y_{0,1}^2 \rangle^{1/2} \simeq \SI{64}{\pico\meter}$ as calculated in \cite{SM}. For small, thermally driven phase shifts, $\Phi_p(L)$ likewise oscillates predominantly at $f_{p}^{y,A}$ with rms amplitude linearly proportion to $y$ displacement, $\langle \Phi_p^2 \rangle^{1/2} \propto \langle Y_{0,p}^2 \rangle^{1/2}$. Far from a band edge, both $\Phi_p$ and $Y_{0,p}$ should be Gaussian random variables, with for example, probability density $P(\Phi_p) = {e^{-\Phi_p^2/2\sigma_p^2}}/\sqrt{2\sigma_p}$.

\subsection{Measurements of phase and amplitude modulation}

\begin{figure}[t]
\centering
\includegraphics[width=1.0\linewidth]{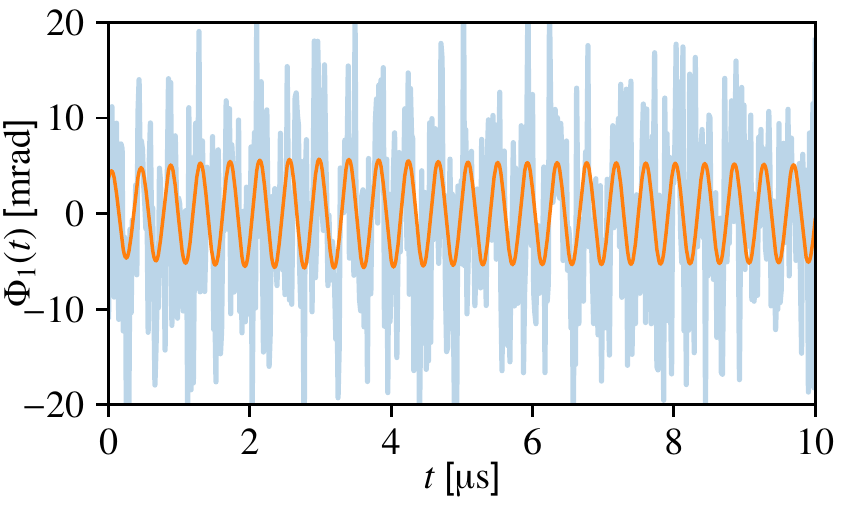}
\caption{Single-shot relative phase between probe signal and local oscillator field extracted from the balanced homodyne photocurrent. (Blue) The data are recorded on a digital oscilloscope with a \SI{62.5}{\mega\hertz} sampling rate. (Orange) Band-pass filtered data with high-cut and low-cut frequencies at $f_1\pm$\SI{100}{\kilo\hertz}. The optical wavelength of the probe is $895.00$ nm.}
\label{fig:AM_FM}
\end{figure}

Overall, our simple model describes mechanical motion via eigenmodes $Y_p^{A}(x)$ that modifies the dispersion relation for an optical GM, which in turn leads to nonzero phase modulation $\Phi_p$ at frequency $f_{p}$ for $p$ odd eigenmodes and zero phase modulation for $p$ even modes, precisely as observed in Fig.~\ref{fig:spectra}(a) far from the band edge. Here we present measurements to substantiate further this model.

With reference to Fig.~\ref{fig:setup}(d), the balanced homodyne detector enables measurement of an arbitrary phase quadrature by offset of the relative phase $\theta$ between the probe output field $E_{out}$ and the local oscillator field $E_{LO}$ with $\theta$ set by adjusting the voltage of the piezoelectric mirror mount (PZT) shown in Fig.~\ref{fig:setup}(d). Phase or amplitude modulation of the probe field is then unambiguously identified by offset $\theta = \pi/2$ for PM or $\theta =0$ for AM. By calibrating the low-frequency ($f \simeq \SI{80}{\hertz}$) fringe amplitude for the difference current $\Delta i(t)$ of the balanced homodyne signal as a function of $\theta(t)$ and then setting $\theta = \pi/2$ (i.e., at the zero-crossing of the interferometer fringe signal for highest phase sensitivity), we observe periodic variation in $\Delta i(t)$ at $f \simeq \SI{2.384}{\mega\hertz}$, corresponding precisely to the lowest $p=1$ eigenfrequency $f_1^{y,A}$ in the phase $\Phi_1(t)$ imprinted on the probe from propagation through the APCW. Figure \ref{fig:AM_FM} displays an example of a single time trace for fixed $\theta = \pi/2$ clearly evidencing $\Phi_1(t)$ both for broad bandwidth detection and for processing with a digital bandpass filter centered at $f_1^{y,A}$ with $\pm \SI{100}{\kilo\hertz}$ bandpass.

Over a range of probe powers (Fig.~\ref{fig:S2}) and frequencies far from the band edge (Fig.~\ref{fig:S3}), the typical observed rms amplitude of the detected phase modulation at $f_1^{y,A}$ is \SI{4.5(20)e-3}{\radian}. This measured modulation for $\Phi_1(t)$ should be compared to the value predicted from our simple model. The thermally driven $y$ amplitude $Y_{0,p=1}$ is calculated in \cite{SM}, and can be combined with a transduction factor $\xi(\omega)=(\frac{dk_x(\omega)}{dy})$ inferred from band structure calculations to arrive to a predicted rms value for thermally driven phase modulation at frequency $f_1^{y,A}$ of about $\SI{4e-3}{\radian}$  \cite{SM}. In Section $5$ we will address the origin of disparity between measured and modeled phase modulation by way of full numerical simulation for the APCW.

Note that we observe a shift of the mechanical frequency with guided probe power, which allows an inference of the bare mechanical frequency $f_1^0$ in the absence of probe light. Representative data for the power-dependent shift can be found in Fig.~\ref{fig:S1}, which shows a linear decrease with probe power $P$ of $f_1 = f_1^0 + \beta P_{\text{out}}$ with $\beta=\SI{-1.31\pm 0.02}{\hertz\per\micro\watt}$ and $f_1^0 =\SI{2385812\pm 10}{\hertz}$. This shift with probe power is consistent with thermal expansion of the APCW due to absorption of probe power.

\subsection{Missing modes}

There remains the question of `missing modes'. If indeed the dominant spectral peaks in Fig. \ref{fig:spectra} are associated with the eigenfunctions $Y_p^{A}$, what has become of the other three sets of eigenfunctions $Y_p^{S}, Z_p^{A},Z_p^{S}$? The answer provided by our simple model of mechanical motion modifying the dispersion relation $k_x(\omega)$ is that $Y_p^A$ is unique in producing a large first-order change in $k_x(\omega)$ with displacement.

Figure \ref{fig:modes} reveals that only $Y_p^A$ has distinct geometries for displacements $\pm \delta y$ (i.e., the two nanobeams are more separated for $+\delta y$ and less separated for $-\delta y$) leading to a much larger calculated transduction factor $\xi_{y,A}(\omega)$ for motion along $y$ than $\xi_{z,A}(\omega)$ for motion along $z$. Moreover, far from the band edge, the symmetric modes $Y_p^{S},Z_p^{S}$ have small transduction factors $\xi_{y,S}(\omega),\xi_{zS,}(\omega)$ comparable to those for modes of a single unmodulated nanobeam of the thickness and average width of the APCW. This issue is addressed in quantitative detail in Section $5$ with a full numerical simulation of  optomechanical coupling for the APCW.

\subsection{Optical frequencies near a band edge}

Near the band edge of a PCW, the mapping of mechanical motion to modification of an optical probe has a qualitatively distinct origin from that in the previous section for the dispersive regime of a PCW. For a finite length PCW, there appears a series of optical resonances $\nu_n$ with $n=1,2,3...$ as displayed in Figure \ref{fig:setup}(c). Each optical resonance arises from the condition $\delta k_x(n) = k_{BE}- k_x = n\pi/L$ with $k_{BE}=\pi/a$ at the band edge \cite{hood2016atom,hoodthesis}. The mapping from wave vector $\delta k_x(n)$ to frequency $\nu_n$ involves a nonlinear dispersion relation $\delta k_x(\nu)$ near the band edge, which for our devices takes the form 
\begin{equation}  
\delta k_x(\nu) = \frac{2 \pi}{a} \sqrt{ \frac{(\nu_{\rm BE2}-\nu)(\nu_{\rm BE}-\nu) }{ 4 \zeta^2 - (\nu_{\rm BE2} - \nu_{\rm BE})^2 }}, 
\label{eq:VH}
\end{equation}
where $\nu_{\rm BE}$ ($\nu_{\rm BE2}$) is the lower (upper) band edge frequency, and $\zeta$ is a frequency related to the curvature of the band near the band edge. Validation of this model by measurement and numerical simulation is provided in Refs.~\cite{hood2016atom,hoodthesis}.

For our current investigation, the lower frequency $\nu_{\rm BE}$ for which $\delta k_x = 0$ is the dielectric band edge frequency. We model how displacements of the APCW geometry for the various mechanical eigenmodes illustrated in Fig. \ref{fig:modes} lead to variation of the parameters in Equation \ref{eq:VH}. Specifically, since the resonance condition involves only the effective length of the APCW (i.e., $L = (N-1)a$ with the number of unit cells $N \simeq 150$ and lattice constant $a\simeq \SI{370}{\nano\meter}$), each optical resonance will be taken to have fixed $\delta k_x(n)=n/(N-1)\times k_{BE}$ with then the associated optical frequency $\nu(n)$ changing due to variation of parameters in Eq. \ref{eq:VH} driven by displacements from the mechanical eigenmodes.$^{\dag\dag}$

\footnotetext{\small $^{\dag\dag}$ In this regard, operation in the vicinity of an optical resonance near a band edge of a PCW is analogous to more traditional opto-mechanics, with, for example, Fabry-Perot cavities, for which thermally excited mechanical resonances of a cavity mirror can shift the optical resonances of a high-finesse cavity. The result on a circulating optical field can be phase or amplitude modulation, or even more exotic behavior, including parametric instability \cite{rokhsari05,braginsky01}, which we will briefly discuss in the concluding section.} 

A mapping of changes in device geometry to changes in band edge frequencies is provided in Ref.~\cite{mcclungthesis}. As in the previous subsection, we seek here a qualitative description to understand the complex transduction of mechanical motion to optical modulation in a $3D$ PCW. Quantitative numerical calculations will be described in the next section.

That said, we proceed by way of Table 2.1 and Figure 2.13 in Ref.~\cite{mcclungthesis} to estimate the traditional optomechanical coupling coefficient $G_{\nu}^y$ for $y$ displacements at the $n=1$ optical resonance, $\nu_1$, closest to the dielectric band edge at $\nu_{BE}$. Here, $G_{\nu}^{y}(\nu_1) \equiv 2 y_{zp}\times\frac{d\nu(1)}{dy}$, where we consider change in resonant frequency $\nu_1$ due to $y$ variation of the gap width $g$ as from the simple model in the previous section, and where the factor $2$ arises for the eigenmode $Y^A$ from the displacement $2 \delta y$ for asymmetric $y$ motion of each beam by $\pm \delta y$ and $\mp \delta y$. $y_{zp}=\sqrt{\hbar/2 m_{\text{eff}}^{y}\omega_p} \simeq \SI{14}{\femto\meter}$ is the zero-point amplitude along the chosen coordinate $y$ \cite{SM}, with the effective mass of a 1D string $m_{\text{eff}}^y=m/2$ and the mass $m\simeq\SI{35}{\pico\gram}$ corresponding to that of the APCW section plus half the mass of each taper. By way of the dispersion relation Eq. \ref{eq:VH} and Ref.~\cite{mcclungthesis}, we find that $\left|\frac{d\nu(1)}{dy}\right| \simeq \SI{0.034}{\tera\hertz\per\nano\meter}$, and thus that the optomechanical coupling coefficient $G_{\nu}^{y}(\nu_1)\simeq \SI{900(100)}{\kilo\hertz}$, which is to be compared to the value found in the following section for the full $3D$ geometry.

\vspace{-5mm}
\section{Numerical evaluation of the opto-mechanical coupling rate $G_{\nu}$}

\begin{figure}[t]
\centering
\includegraphics[width=1.0\linewidth]{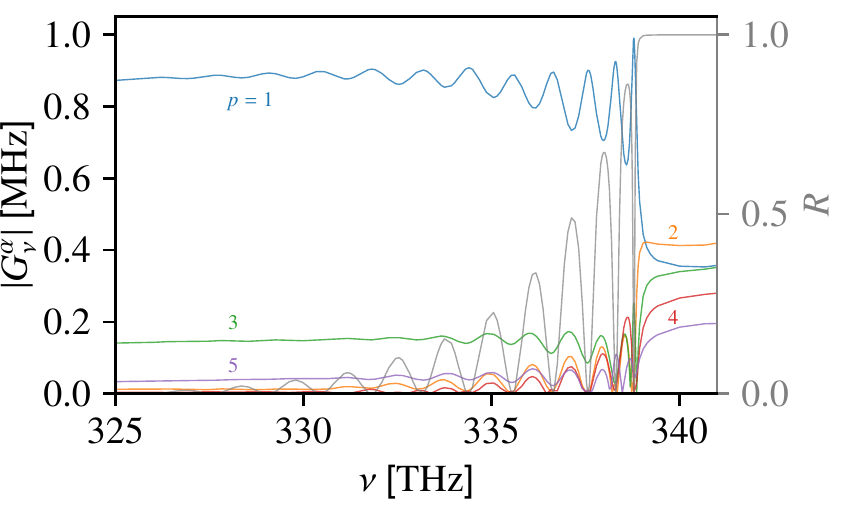}
\caption{Numerically calculated opto-mechanical coupling rate $G_{\nu}^{\alpha}$ for eigenmodes from $p = 1$ to $p = 5$ for the family $Y_p^A$ as functions of optical frequency $\nu$ for a TE guided mode. $\alpha$ is the generalized displacement coordinate defined in \cite{SM}.  Grey curve shows the reflection spectrum for the TE mode of APCW structure. Here, the number of APCW unit cells is $N=150$, the total number of taper cells is $N_t=30$, and the Y-split junction length is $L_Y = \SI{30}{\micro\meter}$. }
\label{fig:coupling}
\end{figure}

In this section, we consider the full APCW structure and evaluate numerically the opto-mechanical coupling rate $G_\nu$ from the waveguide to the band-edge regions.
We first solve for the light field distribution propagating in the structure by launching the TE mode solution of the infinite single nanobeam waveguide section. This also gives reflection and transmission coefficients of the TE electromagnetic mode at both ends of the structure, with the reflection coefficient $R(\nu)$ shown on the right axis of Fig.~\ref{fig:coupling}. We neglect the small imaginary part of the refractive index for SiN as well as losses due to fabrication imperfections. The mechanical eigenmodes are solved for the full structure (i.e., total number of unit cells for APCW $N=150$, total number of taper cells $N_t =30$, Y-split junction length $L_Y = \SI{30}{\micro\meter}$) with clamped ends, taking into account a constant stress distribution which is the steady-state stress field associated to the e-beam written geometry within the sacrificial layer of SiN with initial homogeneous in-plane stress $T$.

Exploring SiN material properties within $10\%$ of the values provided by the wafer manufacturer, the numerically predicted mechanical frequencies are accurate to better than $0.1\%$ with measured frequencies for $E=\SI{250}{\giga\pascal}$, $\rho = \SI{3160}{\kilo\gram\per\cubic\meter}$ and $T=\SI{860}{\mega\pascal}$.

The exact expression for the opto-mechanical coupling rate $G_\nu$ due to displacement shifts of the dielectric boundaries within perturbation theory can be found in \cite{johnson02,SM}. It is given by the product of the mechanical zero-point motion amplitude $\alpha_{zp}$, and the change in optical mode eigenfrequency due the dielectric displacement prescribed by the mechanical mode (generalized coordinate $\alpha$, \cite{SM}), $G_\nu^\alpha = (\partial\nu /\partial\alpha) \alpha_{zp}$.

The values of the coupling rate $G_\nu^\alpha (p)$ are shown in Fig. \ref{fig:coupling} for various eigenmodes $p$ for the family $Y_p^A$ as functions of optical frequency, where the actual eigenmode was approximated by a sine mode shape in Section IV. While the predicted $G_\nu^\alpha$ is largest for such mode family, we report in supplemental Fig.~\ref{fig:S4} the simulated values for all low-frequency modes. The calculation spans from the waveguide regime far below the TE dielectric band edge, to then approaching the band edge, and finally into the band gap itself. The value of $|G_\nu^\alpha|$ reaches up to $\sim \SI{1.0}{\mega\hertz}$ at resonance near the band edge. This is slightly larger than predicted from the simple model in Section IV, which ignored the finite geometry with the Y-junction, tapered cells and narrowing of the physical gap (i.e., infinite APCW).

In contrast to the strains associated with \SI{}{\giga\hertz}-acoustic modes for some optomechanical systems \cite{eichenfield09} that lead to photo-elastic contributions $G_{\text{PE}}$ comparable to those from the dielectric moving boundaries, we find that the $G_{\text{PE}}$ contribution is negligible (by several orders of magnitude) as compared to the dielectric moving boundary contribution for the long-wavelength vibrations under consideration for the APCW, for which the phonon wavelength becomes comparable to the optical wavelength. A measurement of the photo-elastic constant for SiN can be found in \cite{gyger2020}. Also note that $G_{\text{PE}} \propto n^4$, with the ratio of SiN (as here) to Si (as in \cite{eichenfield09}) refractive indices $n_{\text{SiN}}/n_{\text{Si}} = 2/3$.

To validate our numerical calculations, we have reproduced published results for several nanophotonic structures, \cite{eichenfield09j,burek16,Li2015}, as discussed in the Supplemental Material. 

Despite their relatively large effective mass ($\simeq \SI{20}{\pico\gram}$), the low frequency mechanical modes of the APCW achieve mass-frequency products and hence zero-point motion similar to that for 1D structures with microwave phonons coupled to a photonic defect light mode \cite{eichenfield09j}. For comparison of the APCW with 2D structures (as in \cite{tsaturyan17}), the mechanical modes are in the few MHz domain in both cases, but have an effective mass which is two orders of magnitude larger ($\sim\SI{10}{\nano\gram}$) for the $2D$ case.
\begin{figure}[t]
\centering
\includegraphics[width=1.0\linewidth]{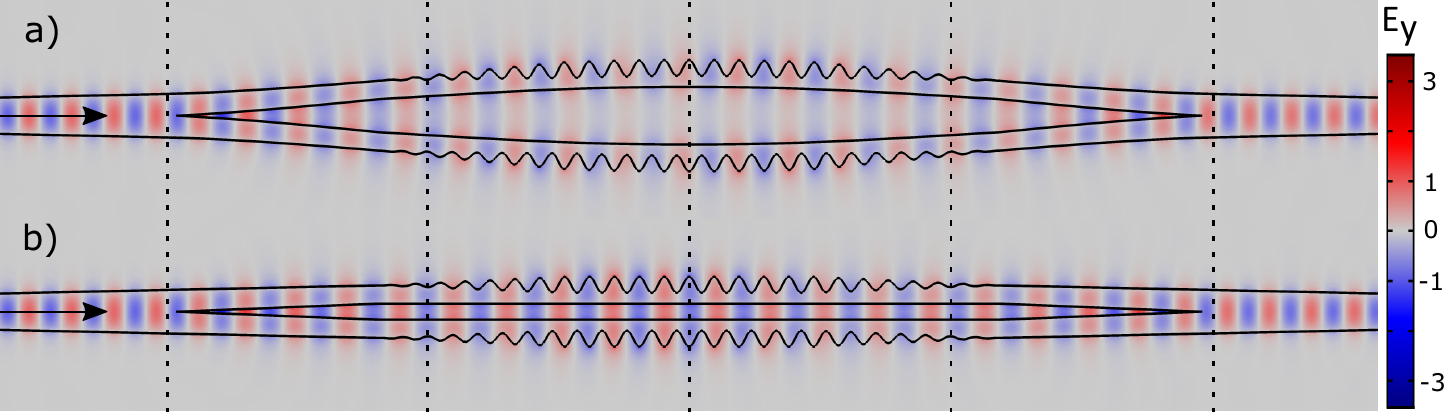}
\caption{Electric field component $E_y$ of the TE guided mode propagating from left to right in the deformed (a) and unperturbed (b) APCW structure, both shown in the $z=0$ mid-plane. The optical frequency is $\SI{320}{\tera\hertz}$. The deformation is prescribed by $Y_1^A$ and here for clarity of illustration is chosen to be an unreasonably large displacement amplitude producing a relative $\pi$ phase shift at the APCW output. $E_y$ is normalized to its maximum strength at single nanobeam input. Here, the number of APCW unit cells is $N=10$, the total number of taper cells is $N_t=15$, and the Y-split junction length is $L_Y = \SI{5}{\micro\meter}$.}
\label{fig:nonperturbative}
\vspace{-10pt}
\end{figure}

Beyond traditional perturbation theory which utilizes the unperturbed optical fields to evaluate $G_\nu^\alpha$, a powerful approach to confirm the transduction mechanism consists in solving Maxwell's equations for the propagation of light
in the deformed dielectric geometry at all phases of the prescribed mechanical eigenmode. Figure~\ref{fig:nonperturbative} illustrates this method, where the deformation of the dielectric produces a relative phase shift on the output light relative to the undeformed case. Owing to the large mismatch between optical $k_x$ and acoustic $q$ wave-vectors, the deformation is quasi-adiabatic. In particular for our very long structure and picometer thermal amplitude, radiation losses into non-guided modes are negligible. With this approach we anticipate weaker phase modulation for $Y^S$, $Z^A$ and $Z^S$ motions to occur at twice their respective eigenfrequency.

\vspace{-5mm}
\section{Conclusion and outlook}

We have reported measurements and models that investigate the low frequency, thermally driven motion of the normal modes of an APCW and the transduction of this motion to the amplitude and phase of weak optical probe beams propagating in a TE guided mode both far from and near to the dielectric bandedge of the APCW. The in-plane antisymmetric mode $Y_p^A(x)$ of the two corrugated nanobeam oscillators dominates the opto-mechanical coupling to TE guided mode light. Simple models describe the basic transduction mechanisms both in the waveguide region far from a band edge as well as in a ``cavity-like'' regime for frequencies near a band edge. 

Beyond simple models, full numerical simulations of the APCW structure have been carried out for quantitative predictions of optomechanical coupling $G_{\nu}$ as in Figure \ref{fig:coupling}. An example is the prospect for detection of zero-point motion $\alpha_{zp}(p=1) \simeq \SI{14.7}{\femto\meter}$. Following the analysis in Ref. \cite{clerk10}, we find probe power $P \simeq \SI{10}{\micro\watt}$ would be sufficient to reach phase sensitivity corresponding to $\alpha_{zp}$ for measurement bandwidth equal to the current linewidth $\gamma_1\sim \SI{24}{\hertz}$ for $Y_1^{A}(\alpha_{zp})$ \textit{if} this mode were cooled to its motional ground state. Moreover, the resulting back-action noise from the probe would correspond to $\alpha_{zp}$, thereby reaching the Standard Quantum Limit for $y$ motion of the APCW at $f_1\simeq \SI{2.4}{\mega\hertz}$.

While the quality factors are modest for the APCW compared to current best literature values, the very small effective mass of the APCW allows for thermo-mechanical force sensitivity at a limit of $\sqrt{S_{FF}}\simeq\SI{143}{\atto\newton\per\sqrt{\hertz}}$. This value is only $\sim 2.6\times$ times larger than that of \cite{tsaturyan17} ($\SI{55}{\atto\newton\per\sqrt{\hertz}}$), namely $S_{FF} = 4\pi m_{\text{eff}} f_1 k_B T/Q$. 

In terms of cooling to the ground state from a room-temperature APCW, the minimum $Q$-frequency product $Q \cdot f = \SI{6e12}{\hertz}$ \cite{wilson09} would require $Q$ values about $26\times$ larger than currently observed. Certainly, many advanced design strategies are available for increasing quality factors for a ``next-generation'' of $1D$ PCWs  \cite{tsaturyan17}. In addition, low GM powers lead to strong pondermotive forces within the gap of the APCW that could potentially be harnessed to increase mechanical quality factors by $\sim 50 \times$ by way of ``optical springs'' \cite{ni12}.
Beyond the focus of this article, we can excite selectively the observed mechanical
modes with amplitude-modulated guided light at the specific observed frequencies. In fact, we also observe driving of the
mechanical resonances with the external optical conveyor belt described in \cite{Burgers19}.

As for optical cooling of the APCW, our initial measurements related to opto-mechanics in a nonlinear regime suggest that efficient cooling might be achieved by operating near a bandedge. For example, as illustrated by Fig.~\ref{fig:S5} in \cite{SM}, we observe low-power bistable behavior marked with strong self-oscillation (near radian-phase modulation amplitude) for continuous GM power thresholds below \SI{100}{\micro\watt}. The large scale oscillations could originate from thermal effects of the APCW due to the GM light, which we are investigating. Alternatively, the bistable behavior and self-induced oscillates might arise from optical spring effects as described in Ref.~\cite{rokhsari05}. A double-well potential with two stable local minima can be developed when the GM power is sufficiently high \cite{aspelmeyer14}. The detailed mechanism of the large scale oscillations is beyond the scope of this paper, and will be investigated in our subsequent experiments. Related instabilities for blue cavity detunings are a hallmark of cooling for red detunings in conventional opto-mechanics in optical cavities \cite{kippenberg07}.

The observations on mechanical modes of the APCW reported here are also important for assessing deleterious heating mechanisms for combining atom trapping in the vicinity of nano-photonic structures \cite{zoubi16}. While the symmetric modes lead to negligible modulations of the guided light as compared to $Y^A$ motion, the guided light intensity distribution still follows the motion of the APCW structure in the laboratory frame. A simple estimate of heating limited trap lifetime due to trap-potential pointing instability can be obtained from the thermal position instability of $\sqrt{S_{yy}}\sim\SI{3.8}{\pico\meter\per\sqrt{\hertz}}$ at $f_1$, with $S_{yy} = 2k_B T Q/m_{\text{eff}}\omega_1^3$ \cite{clerk10}. This noise level corresponds to an energy-doubling time $\tau$ \cite{savard97} of order \SI{1}{\milli\second}, at atom trap frequency $f_1$. We are working on further simulations of heating rates with the complex motion of these dielectric structures for cold atom traps. Implementing feedback cooling with guided light could also mitigate limitations from operation at room-temperature \cite{zhao12}.

\begin{figure}[t]
\centering
\vspace{10pt}
\includegraphics[width=\linewidth]{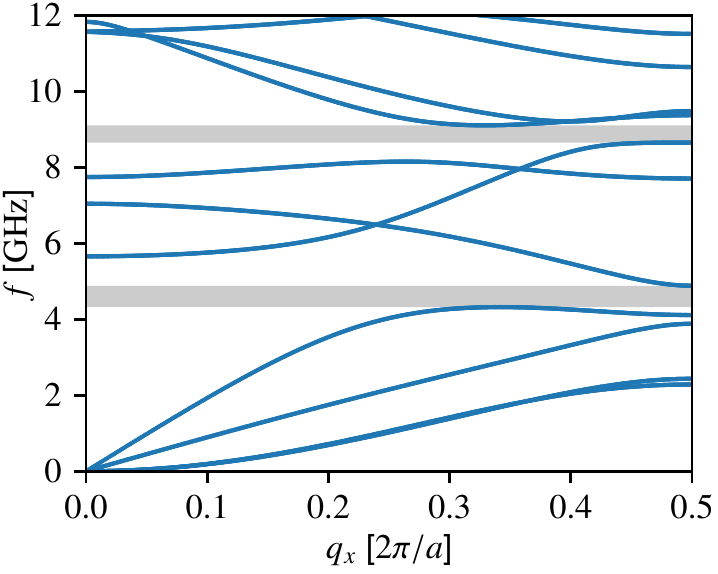}
\caption{Phononic band diagram of the infinite APCW structure, with the acoustic wavevector component $q_x$ spanning the irreducible Brillouin zone. The gray-shaded areas represent band gaps, with opportunity for flat bands and band-gap physics with frequencies tuned to the hyperfine ground state frequency of alkali atoms.}
\label{fig:GHz}
\vspace{-10pt}
\end{figure}

Although we have concentrated on low-frequency eignmodes of the APCW in the MHz regime, we have also investigated eigenmodes in the GHz regime that are of interest for many of the topics addressed here. As illustrated in Fig. \ref{fig:GHz}, the corrugated structure of the APCW can lead to \textit{phononic} band gaps in the GHz acoustic domain. The possibilities for band-gap engineering for both photons and phonons \cite{eichenfield09j} for application to atomic physics (e.g., for coupling mechanics to both Zeeman and hyperfine atomic states) represent an exciting frontier beyond the work reported here. One example to note is that the curvature of phonon bands can strongly enhance heating rates for atom traps \cite{hummer19}, which might offer new possibilities for engineering better atom traps in PCWs for atomic physics.

\vspace{-5mm}
\begin{acknowledgments}
\vspace{-5pt}
The authors acknowledge sustained and important interactions with A. P. Burgers, L. S. Peng, and S.-P. Yu, who fabricated the nano-photonic structures used for this research. JBB acknowledges enlightening discussions with Y. Tsaturyan. HJK acknowledges funding from the Office of Naval Research (ONR) Grant \#N00014-16-1-2399, the ONR MURI Quantum Opto-Mechanics with Atoms and Nanostructured Diamond Grant \#N00014-15-1-2761, the Air Force Office of Scientific Research MURI Photonic Quantum Matter Grant \#FA9550-16-1-0323, and the National Science Foundation (NSF) Grant \#PHY-1205729.
\end{acknowledgments}

\vfill

\newpage

\appendix
\section{Supporting figures}

\begin{figure}[hbt]
\centering
\includegraphics[width=\linewidth]{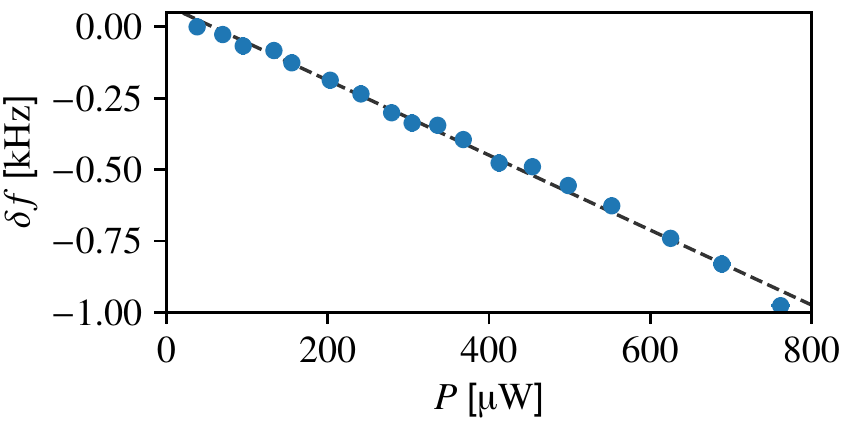}
\caption{Frequency shift $\delta f$ of the lowest frequency peak at $f_1 \simeq \SI{2.4}{\mega\hertz}$ (see Fig.~\ref{fig:spectra}(a) of main text) as a function of transmitted probe power $P_{out}$. (Dashed line) A linear fit of $f_1 = f_1^0 + \beta P_{\text{out}}$ with $\beta=\SI{-1.31\pm 0.02}{\hertz\per\micro\watt}$ and $f_1^0 =\SI{2385812\pm 10}{\hertz}$.}
\label{fig:S1}
\end{figure}

\begin{figure}[hbt]
\centering
\includegraphics[width=\linewidth]{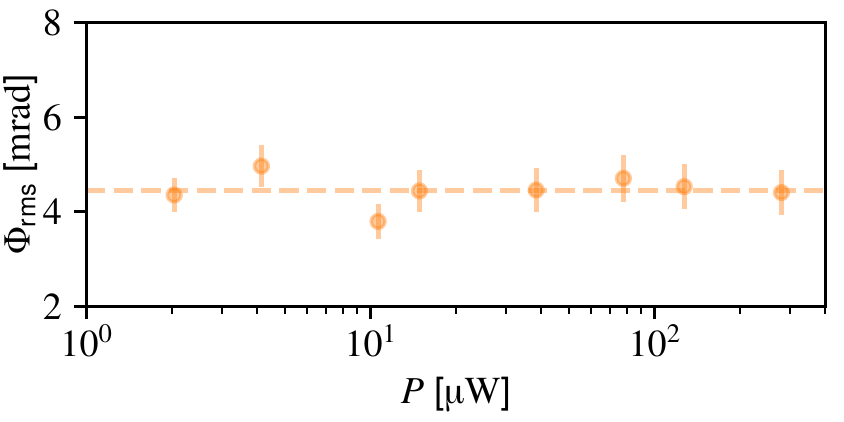}
\caption{Measured RMS phase modulation from thermal excitation at $\SI{300}{\kelvin}$ as a function of output probe power. Each point is an average value over 100 records, with each record obtained in the manner of Fig.~\ref{fig:AM_FM} of main text.}
\label{fig:S2}
\end{figure}

\begin{figure}[hbt]
\centering
\includegraphics[width=\linewidth]{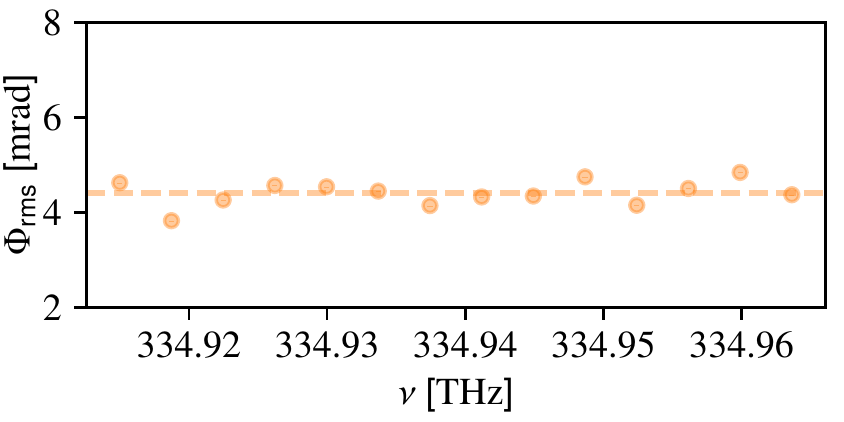}
\caption{Measured RMS phase modulation from thermal excitation at $\SI{300}{\kelvin}$ as a function of probe GM optical frequency for the range shown in red in Fig.~\ref{fig:setup}(c) (main text) (from $\sim (\nu_a-\SI{50}{\giga\hertz})$ to $\sim \nu_a $). Each point is an average value over 100 records, with each record obtained in the manner of Fig.~\ref{fig:AM_FM} of main text.}
\label{fig:S3}
\end{figure}

\begin{figure}[hbt]
\centering
\includegraphics[width=\linewidth]{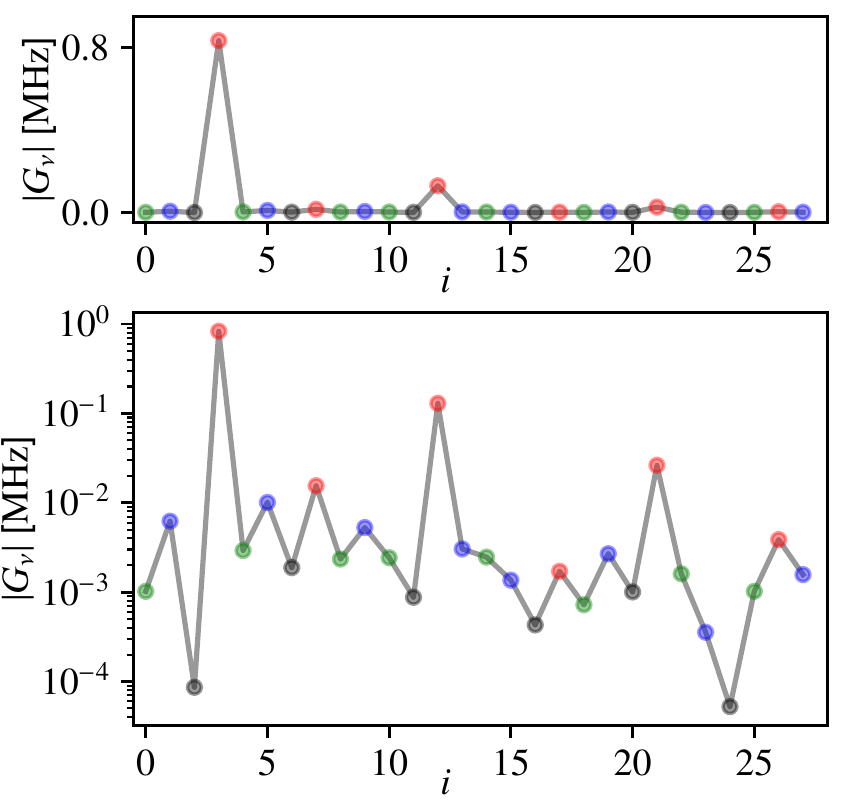}
\caption{Calculated opto-mechanical coupling rate $G_\nu^\alpha$ for the lowest mechanical modes of all families evaluated at $\nu=\SI{320}{\tera\hertz}$ and sorted by ascending mechanical eigenfrequencies with index $i$. (green) Symmetric Z eigenmodes; (blue) Symmetric Y; (red) Anti-symmetric Y; (black) Anti-symmetric Z. Upper panel: modulus of $G_\nu^\alpha$ on a linear scale. Lower panel: modulus of $G_\nu^\alpha$ on a logarithmic scale. Note that for anti-symmetric Y modes (red) the predicted suppression of the quasi-even harmonics agrees with our measurement.}
\label{fig:S4}
\end{figure}

\begin{figure}[hbt]
\centering
\includegraphics[width=0.95\linewidth]{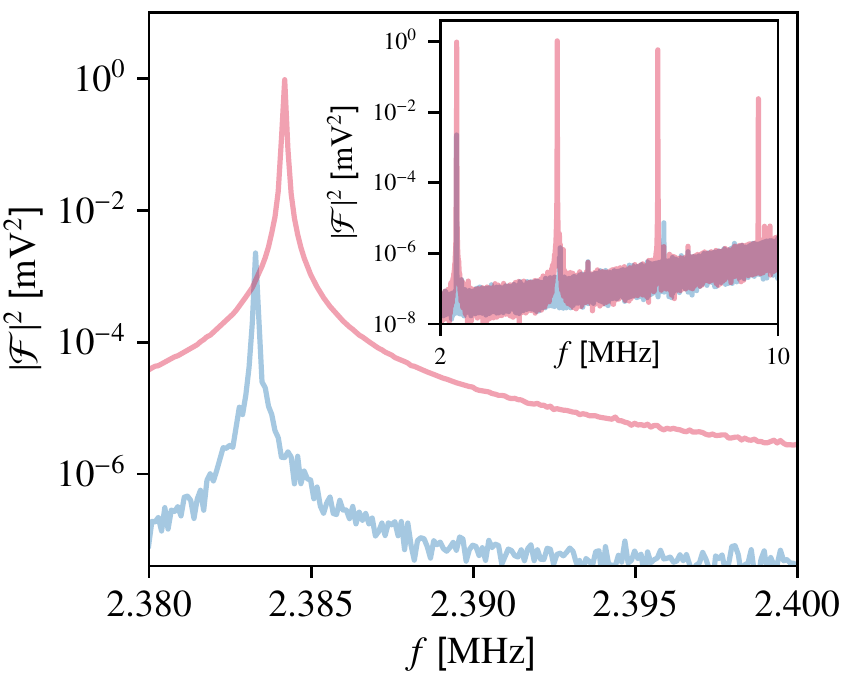}
\caption{Bi-stable modulation observed near the band edge marked by very strong self-oscillation (also seen with harmonics shown in inset). The observed regeneration  time scale (Ref.~\cite{rokhsari05}) is $\sim\SI{3}{seconds}$.  (Optical frequency $\nu=\SI{342.76}{\tera\hertz}$).}
\label{fig:S5}
\end{figure}

Figures \ref{fig:S1}, \ref{fig:S2}, \ref{fig:S3}, \ref{fig:S4}, and \ref{fig:S5}  provide supporting information for the measurements and numerical models presented in the main manuscript.

\vspace{-5mm}
\appendix
\section{Zero-point and thermal amplitudes}

Following existing definitions for optomechanical crystal devices as in \cite{eichenfield09j}, the amplitude of the displacement vector field $\textbf{u}$ is parametrized by a generalized coordinate $\alpha$ such that $\textbf{u} = \alpha\hat{u}$ where $\hat{u}(\textbf{r}) \equiv \textbf{u}(\textbf{r})/\text{max}(|\textbf{u}(\textbf{r})|) $ is the unit-normalized displacement vector field with $\text{max}(|\hat{u}(\textbf{r})|) = 1$. From our full Comsol simulations, we find the maximum gap width is $2\times\alpha$ for $Y^A_p$ motion.

The amplitude of zero-point motion of mode $p$ is \cite{aspelmeyer14,aspelmeyer2014cavity})
\begin{align*}
\alpha_{zp}(p) = \sqrt{\frac{\hbar}{2m_{\text{eff}}^{\alpha}\omega_p}},
\end{align*}
where $\omega_p = 2\pi f_p$ is the mechanical angular frequency, and $m_{\text{eff}}^{\alpha}$ is the effective mass with the associated definition 
\begin{align*}
m_{\text{eff}}^{\alpha} = \frac{\int \rho(\textbf{r}) |\textbf{u}(\textbf{r})|^2 \text{dV} }{\text{max}( |\textbf{u}(\textbf{r})|)^2}, \quad \alpha \equiv \text{max}( |\textbf{u}(\textbf{r})|),
\end{align*}
where $\rho$ is the scalar mass density of the material.

From the numerical solution of the fundamental ($p=1$) mechanical mode, $m_{\text{eff}}^{\alpha}\simeq \SI{16.3}{\pico\gram}$ and $\alpha_{zp}(1) \simeq \SI{14.7}{\femto\meter}$. Note that the bulk mass of the simulated structure is \SI{45.4}{\pico\gram}.

At room temperature $T=\SI{300}{\kelvin}$, $\hbar\omega_1 \ll k_B T$, hence the mean thermal phonon number $\bar{n}_{\text{th}}(\omega_1) = 1/( \text{Exp}(\hbar \omega_1 / k_B T)-1) \simeq k_B T/\hbar\omega_1 =  \SI{2.6e6}{}$. This gives an rms thermal amplitude $\alpha_{\text{th}} \simeq \alpha_{zp}\sqrt{ 2 \bar{n}_{\text{th}}} = \SI{33.5}{\pico\meter}$.

\smallskip
\textbf{Equipartition theorem}:
The same result is obtained from the classical equipartition theorem. The rms amplitude $\langle \alpha^2_p \rangle^{1/2}$ of mode $p$ in thermal equilibrium at temperature $T$ is
\begin{align*}
\langle \alpha^2_p \rangle^{1/2}= \sqrt{\frac{k_B T}{m_{\text{eff}}^{\alpha}\omega_p^2}},\quad \langle \alpha_{p=1}^{2} \rangle^{1/2} \simeq \SI{33.4}{\pico\meter}.
\end{align*}

\smallskip
\textbf{Simple model}:
For the simple model, we consider the material mass $m\simeq\SI{35}{\pico\gram}$ corresponding to the full APCW section plus half the mass of each taper. With the normalization of the simplified eigenmode $Y_{p=1}^{A}(x)$ in section IV, the effective mass associated to the amplitude $Y_{0,p}$ is $m_{\text{eff}}^{Y} = m/8$, where a factor $1/2$ arises from the sinusoidal mode shape function (1D string) and a factor $1/4$ from $Y_{0,p}$ being the maximum separation between the two nano-beams. When excited by Langevin thermal forces, the mechanical mode $Y_{p=1}^{A}(x)$ will oscillate along $y$ at frequency $f_{p=1}^{y,A}$ with rms amplitude 
\begin{align*}
\langle Y_{0,p=1}^{2} \rangle^{1/2} = \sqrt{\frac{k_B T}{m_{\text{eff}}^{Y}\omega_1^2}} \simeq \SI{64.4}{\pico\meter},
\end{align*}
with $\langle Y_{0,p=1}^{2} \rangle^{1/2}\simeq 2\times \langle \alpha^2_{p=1} \rangle^{1/2}$ and $m_{\text{eff}}^{Y} \simeq m_{\text{eff}}^{\alpha}/4$.

Estimates for the transduction of motion to modulation for the simplified eigenmode $Y_{p=1}^{A}(x)$ are derived from Fig.~2.13 in \cite{mcclungthesis} and Eq. (2) in Section IV (D).

\smallskip
\appendix
\section{Opto-mechanical coupling $G_{\nu}$, Data processing, and Validation of simulations}

\smallskip
\textbf{Opto-mechanical coupling rate}: 
$G_\nu^\alpha = (\partial\nu /\partial\alpha) \alpha_{zp}$, with (Ref.~\cite{johnson02})
\begin{align*}
    \frac{\partial \nu}{\partial \alpha} = -\frac{\nu_0}{2} \frac{\iint (\textbf{u}(\textbf{r})\cdot \textbf{n})(\Delta\epsilon |\textbf{E}_{\parallel}(\textbf{r})|^{2}-\Delta\epsilon^{-1}|\textbf{D}_{\perp}(\textbf{r})|^{2}) \text{dS}}{\text{max}( |\textbf{u}(\textbf{r})|) \int \epsilon |\textbf{E}(\textbf{r})|^2 \text{dV}}
\end{align*}
where $\textbf{u}$ is the displacement vector field, $\textbf{n}$ is the unit vector normal to the surface of the dielectric structure, $E$ and $D$ are the unperturbed electric and electric displacement fields (with components parallel or perpendicular to the local surface), $\Delta \epsilon = \epsilon_{\text{SiN}} - \epsilon_{\text{vac}}$ and $\Delta\epsilon^{-1}=1/\epsilon_{\text{SiN}} - 1/\epsilon_{\text{vac}}$.

\smallskip
\smallskip
\textbf{Data filter}: For the measurement reported in Fig.~\ref{fig:AM_FM}, the photocurrent signal is sampled every $\SI{16}{\nano\second}$ with a precision digital oscilloscope. The data are further processed with a fourth-order Butterworth bandpass filter with high-cut and low-cut frequencies $f_1\pm\SI{100}{\kilo\hertz}$.

\smallskip
\smallskip
\textbf{Validation of numerical simulations}:
To validate the numerical predictions for our structure, we find excellent agreement with simulations performed on nanophotonic structures published from other research groups. (FEM simulations are performed with Comsol Multiphysics 5.4.). For instance, for the diamond crystal cavity with triangular beam cross-section in \cite{burek16}, we find $\{g_{MB},g_{PE},g_{\text{tot}}\} = 2\pi\times \{60.1,76.6,136.7\}\si{\kilo\hertz}$ for the flapping mode with $\{f,m_{\text{eff}}^{\alpha}\} = \{\SI{5.99}{\giga\hertz},\SI{144.4}{\femto\gram}\}$; $\{g_{MB},g_{PE},g_{\text{tot}}\} = 2\pi\times \{42.7,187.3,230\}\si{\kilo\hertz}$ for the swelling mode with $\{f,m_{\text{eff}}^{\alpha}\} = \{\SI{8.80}{\giga\hertz},\SI{187.6}{\femto\gram}\}$.

\newpage

\bibliography{main}

\end{document}